\documentclass{fuin}

\usepackage[dvips]{epsfig}

\title{Chaotic Dynamics in Iterated Map Neural Networks with 
Piecewise Linear Activation Function \footnote{I would like to thank
Prof. S. K. Pal (MIU, ISI) for his constant encouragement.}} 

\author{Sitabhra Sinha\\
       Department of Physics\\
       Indian Institute of Science\\                     
       Bangalore - 560 012, India\\  
       sitabhra@physics.iisc.ernet.in
       }
\date{}

\begin{document}
\setlength{\topmargin}{1cm}

\maketitle

\thispagestyle{empty}

\begin{abstract}
The paper examines the discrete-time dynamics of neuron models 
(of excitatory and inhibitory types) with piecewise  
linear activation functions, which are 
connected in a network. The properties of a pair of neurons 
(one excitatory and the other inhibitory) 
connected with each other, is studied in detail. Even such a simple 
system shows a rich variety of behavior, including high-period oscillations 
and chaos. Border-collision bifurcations and multifractal fragmentation 
of the phase space is also observed for a range of parameter values. 
Extension of the model to a larger number of neurons is suggested under 
certain restrictive assumptions, which makes the resultant network dynamics 
effectively one-dimensional. Possible applications of the network for 
information processing are outlined. These include using the network for 
auto-association, pattern classification, nonlinear function approximation 
and periodic sequence generation. 
\end{abstract}

\section{Introduction}
\markboth{S. Sinha}{Chaos in neural networks} 

Since the development of the electronic computer in the 1940s, the serial 
processing computational paradigm has successfully held sway. It has 
developed to the point where it is now ubiquitous. However, there are many 
tasks which are yet to be successfully tackled computationally. A case in 
point is the multifarious activities that the human brain performs regularly, 
including pattern recognition, associative recall, etc. which is extremely 
difficult, if not impossible to do using traditional computation. 
 
This problem has led to the development of non-standard techniques to 
tackle situations at which biological information processing systems 
excel. One of the more successful of such developments aims at 
``reverse-engineering'' the biological apparatus itself 
to find out why and how it works. The field of neural network models 
has grown up on the premise that the massively parallel distributed 
processing and connectionist structure observed in the brain is the key 
behind its superior performance. By implementing these features in the 
design of a new class of architectures and algorithms, it is hoped that 
machines will approach human-like ability in handling real-world situations. 
 
Extremely simplified models of neurons connected in a network  
via suitable connection weights were known to implement various logical 
functions since the 1940s. The subject received fresh impetus a decade and 
half ago due to some breakthroughs. These were however restricted to 
networks subject to convergent dynamics. Such systems tend to a 
time-independent solution (a ``fixed-point'' attractor) after starting  
off from some initial condition. On the other hand, the brain never 
settles down to a steady state but appears to exhibit a rich variety of 
non-periodic behavior. 
 
The development of nonlinear dynamical systems theory - in particular, the 
discovery of ``deterministic chaos'' in extremely simple systems - has 
furnished the theoretical tools necessary for analyzing non-convergent 
network dynamics. Several neurobiological studies have in fact indicated 
the presence of chaotic dynamics in the brain \cite{freeman92} and its 
possible role in biological information processing. Thus, the 
ability to design networks with aperiodic behavior 
promises to add a new dimension to our understanding of how the brain works. 
 
Several efforts in designing and applying chaotic neural networks have 
already appeared 
\cite{xwang91} - \nocite{inoue92} \nocite{tsuda} \nocite{andrey96} 
\nocite{ishii96} \nocite{sinha96} \cite{lwang96}. 
(For further references see \cite{tsuda}). In the present work, a 
particularly simple model whose state is updated after discrete time 
intervals is studied. 
The piecewise linear nature of the model neuron 
used, not only makes detailed theoretical analysis possible, 
but also enables an intuitive understanding of the dynamics, at least 
for a small number of connected elements. This makes it easier to 
extrapolate to larger networks and suggest possible applications. 
The proposed model is also particularly suitable for hardware 
implementation using operational amplifiers (owing to their piecewise 
linear characteristics). 
 
The rest of the paper is organized as follows. 
The basic features of the piecewise linear neural model used is described 
in section 2, along with the biological motivation for such a model. The 
next section is devoted to analyzing the dynamics of a pair of 
excitatory and inhibitory neurons, with self- and inter-connections. 
This simple system shows a wide range of behavior including periodic 
oscillations, chaos and border-collision bifurcations \cite{nusse95}. 
Section 4 extends the model to larger networks under certain 
restrictive conditions. This is followed by a discussion of the possible 
application of the model to various information processing tasks, such as 
associative memory and nonlinear function approximation. 
The rich dynamics of the system allows it to respond to specific 
inputs with periodic or aperiodic responses (in contrast 
to a time-independent 
constant output, as in convergent networks) and also to act as a {\it 
central pattern generator}. We conclude with a short discussion 
on possible directions for further studies. 

\section{Piecewise Linear Neuron}

Let $u_n$ denote the activation state of a model neuron at the $n-$th time 
interval. If $u_n=1$, the neuron is considered to be active (firing), and 
if $u_n=0$, it is quiescent. Then, if $v_n$ is the input 
to the neuron at the $n$th instant, the discrete-time neural dynamics 
is described by the equation 
\begin{equation} 
u_{n} = {\cal F} (v_{n}), 
\end{equation} 
assuming there to be no effects of delay. 
The input $v_n$ is the weighted sum of the activation states at the $(n-1)$-th 
instant of all other 
neurons connected to the neuron under consideration, together with 
external stimulus (if any). 
The form of ${\cal F}$ is decided by 
the input-output behavior of the neuron. Usually, it is taken to be the 
{\em Heaviside step function}, i.e.,  
\begin{equation} 
\begin{array}{lll} 
{\cal F} (z) & = & 1, ~~{\rm if}~~z>~\mu, \\ 
            & = & 0, ~~{\rm otherwise}, 
\end{array} 
\end{equation} 
where $\mu$ is known as the {\it threshold}. 
 
If the {\it mean firing rate}, i.e., the activation  
state averaged over a time interval, is taken as the dynamical 
variable, then a continuous state space 
is available to the system. If $X_n$ be the mean firing rate at the $n-$th 
time interval, then 
\begin{equation} 
X_{n+1}~=~F_{\mu} (\Sigma_{j} W_{j} X^j_n + I_n). 
\end{equation} 
Here, $F$ is known as the {\it activation function} and $\mu$ is the 
parameter associated with it. The first term of the argument represents 
the weighted sum of inputs from all neurons connected with the one under 
study. $W_{j}$ is the weightage for the connection to the $j$th neuron. 
$I_n$ represents the external stimulus at the $n$th instant. 
 
Considering the detailed biology, there are two transforms occurring at the 
threshold element. At the input end, the impulse frequency coded information 
is transformed into the amplitude modulation of the neural current. For 
single neurons, this pulse-wave transfer function is linear over a small 
region, with nonlinear saturation at both extremities. At the 
output end, the current amplitude is converted back to impulse frequency. The 
wave-pulse transfer-function for single neurons is zero below a threshold, 
then 
rises linearly upto a maximum value. Beyond this maximum, the output 
falls to zero due to ``cathodal block''. 
These relations are time-dependent. For example, the slope of the wave-pulse 
transfer function decreases with time when subjected to sustained activation 
(``adaptation'') \cite{freeman92}. 
 
The net transformation of a input by a neuron is therefore given by the 
combined action of the two transfer-functions. Let us approximate 
the nonlinear 
pulse-wave transfer function $F_1$ with a piecewise linear function, such that 
\begin{equation} 
\begin{array}{lll} 
F_1 (z) & = & -c, ~~{\rm if}~~z<~-c/m,\\ 
        & = & m~z, ~~{\rm if}~~-c/m~ \leq z \leq ~1/m,\\ 
		& = & 1, ~~{\rm if}~~z>~1/m. 
\end{array} 
\end{equation} 
The wave-pulse transfer function $F_2$ is represented as 
\begin{equation} 
\begin{array}{lll} 
F_2 (z) & = & 0, ~~{\rm if}~~z<~t,\\ 
        & = & m^{\prime}~(z-t), ~~{\rm if}~~t~ \leq  
         z \leq ~t+(1/m^{\prime}),\\ 
        & = & 0, ~~{\rm if}~~z>~t+(1/m^{\prime}), 
\end{array} 
\end{equation} 
where $m, m^{\prime}$ are the slopes of $F_1, F_2$ respectively, $c$ is the 
inhibitory saturation value and $t$ represents a threshold value. 
 
It is easily seen that the combined effect of the two gives rise to the 
resultant transfer function, $G$, defined as 
\begin{equation} 
\begin{array}{lll} 
G (z) & = & 0, ~~{\rm if}~~z<~t,\\ 
        & = & m~m^{\prime}~(z-t), ~~{\rm if}~~t~ \leq z \leq ~t+(1/m),\\ 
        & = & m^{\prime}~(1-t), ~~{\rm if}~~t+(1/m)~ <  
         z \leq ~t+(1/m^{\prime}),\\ 
	& = & 0, ~~{\rm if}~~z>~t+(1/m^{\prime}). 
\end{array} 
\end{equation} 
In the present work we will assume that $m^{\prime}~(1-t) <<  
t+(1/m^{\prime}) $. This condition ensures that the operating 
region of the neuron 
does not go into the ``cathodal block'' zone. This allows 
us to work with the following simplified 
neural activation function (upon rescaling) throughout the rest of the paper: 
\begin{equation} 
\begin{array}{lll} 
F_a (z) & = & 0, ~~{\rm if}~~z<~t,\\ 
        & = & a~(z-t), ~~{\rm if}~~t~ \leq z \leq ~t+(1/a),\\ 
        & = & 1, ~~{\rm if}~~z>~t+(1/a), 
\end{array} 
\end{equation} 
where $a~(>0)$ is called the {\em gain parameter} of the function (Fig. 1). 
For infinite gain ($a \rightarrow \infty$), the activation function reverts 
to the hard-limiting Heaviside step function.  
\begin{figure}[htbp]
  \vspace{0.2cm}
    \begin{center}
    \epsfig{file = 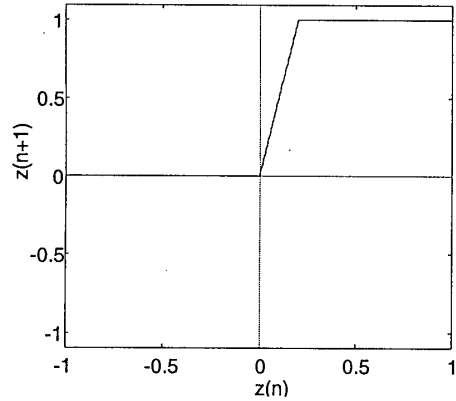, height=6cm}
\caption{The piecewise linear activation function $F$ for a single 
neuron (gain parameter, $a$ = 5).
\label{fuinfig1} }
\end{center}
\end{figure}

Note that, 
if neural populations had been considered, instead of single neurons, then 
sigmoidal activation functions, e.g., 
$$ 
F_{\mu}(z) = 1/(1+e^{- \mu z}), 
$$ 
would have been the appropriate choice. By varying $\mu$, transfer functions 
with different slopes would be obtained. 
 
Although, in the present study, the gain parameter, $a$, 
of the transfer function is 
considered constant, in general it will be a time-varying function of the 
activation state, decreasing under constant external stimulation until the 
neuron goes into a quiescent state. The threshold $t$ is also a dynamic 
parameter, changing as a result of external stimulation. 
We have also assumed that the neuron state at the $n$th instant is a function 
of the state value at the previous instant only. 
Introducing delay effects into the model, such that, 
$$ X_{n+1} = F ( X_n, X_{n - 1}, \ldots, X_{n - \tau} ),$$ 
might lead to novel behavior. 
 
\section{Excitatory-Inhibitory Pair Dynamics} 

Having established the response properties of single neurons, we can now 
study the dynamics when they are connected. It is observed that, 
even connecting only an excitatory and an inhibitory neuron with each 
other leads to a rich variety of behavior, including high period oscillations 
and chaos. The continuous-time dynamics of pairwise connected 
excitatory-inhibitory neural populations (with sigmoidal nonlinearity) 
have been studied before \cite{cowan}. 
However, in the present case, the resultant system is updated in 
discrete-time intervals and the 
dynamics is governed by piecewise linear mappings. This makes chaotic 
behavior possible in the proposed neural network model. 
Chaotic activity has been previously observed in 
piecewise linear systems, for both continuous-time \cite{schulmann} as well as 
discrete-time evolution \cite{nusse92} of the system.  
 
If $X$ and $Y$ be the mean firing rates of the excitatory and inhibitory 
neurons, respectively, then their time evolution is given by the 
coupled difference equations: 
\begin{equation} 
X_{n+1} = F_a ( W_{xx} X_n - W_{xy} Y_n ), 
\end{equation} 
$$ 
Y_{n+1} = F_b ( W_{yx} X_n - W_{yy} Y_n ). 
$$ 
\begin{figure}[htbp]
   \vspace{0.5cm}
      \begin{center}
      \epsfig{file = 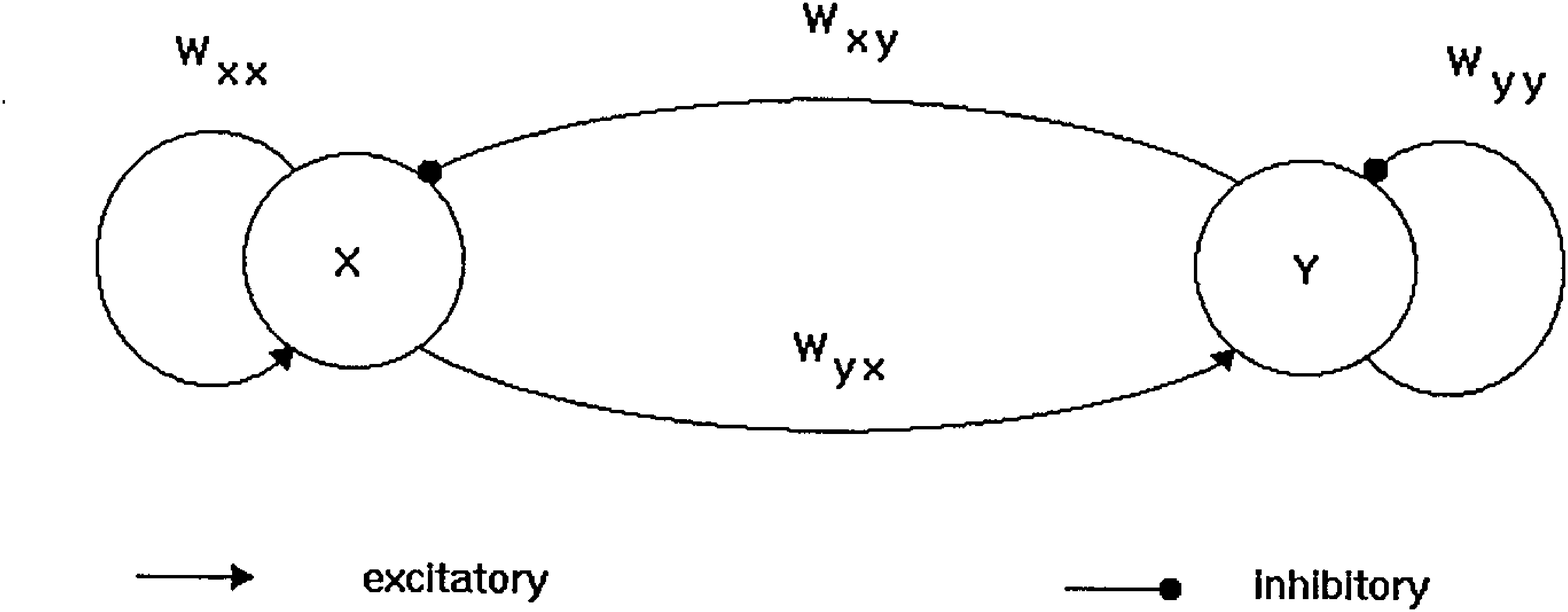, height=5cm}
\caption{ The pair of excitatory ($x$) and inhibitory ($y$) neurons. 
The arrows and circles represent excitatory and inhibitory synapses, 
respectively. 
\label{fuinfig2} }
\end{center}
\end{figure}

The network connections are shown in Fig. 2. Without loss of generality, 
the connection weightages $W_{xx}$ and $W_{yx}$ can be absorbed 
into the gain parameters $a$ and $b$ and the correspondingly rescaled 
remaining connection weightages, $W_{xy}$ and $W_{yy}$, are labeled $k$ 
and $k^{\prime}$ respectively. For convenience, a transformed set of 
variables, $Z_n=X_n - k~Y_n$ and $Z^{\prime}_n=X_n - k^{\prime}~Y_n$, is 
used. The dynamics is now given by 
\begin{equation} 
Z_{n+1} = F_a (Z_n) - k~F_b (Z^{\prime}_n), 
\end{equation} 
$$ 
Z^{\prime}_{n+1}= F_a (Z_n) - k^{\prime}~F_b (Z^{\prime}_n). 
$$ 
Note that, if $k = k^{\prime}$, the two-dimensional dynamics is reduced to 
an effective one-dimensional one, simplifying the analysis. We will now 
examine the cases: $(i) k = k^{\prime} = 1$, $(ii) k = k^{\prime} \neq 1$, 
and $(iii) k \neq k^{\prime}$, in detail. Unless mentioned otherwise, 
the threshold $t$ will be taken as 0 (a non-zero value of $t$ introduces an 
affine transformation to $F$). 
 
\subsection{$k = k^{\prime} = 1$}

This represents the condition when the connection weights $W_{xy}=W_{xx}$ 
and $W_{yy}=W_{yx}$, $(a>b)$. 
The dynamics is that of an {\it asymmetric tent map} (Fig. 3(a)): 
\begin{equation} 
\begin{array}{lll} 
Z_{n+1} & = & (a~-~b)~Z_n, ~~{\rm if}~~0~\leq Z_n \leq~1/a,\\ 
        & = & 1~-~b Z_n, ~~{\rm if}~~~ 1/a~< Z_n \leq~1/b,\\ 
        & = & 0, ~~{\rm otherwise}. 
\end{array} 
\end{equation} 
\begin{figure}[htbp]
  \vspace{0.25cm}
    \begin{center}
      \epsfig{file = 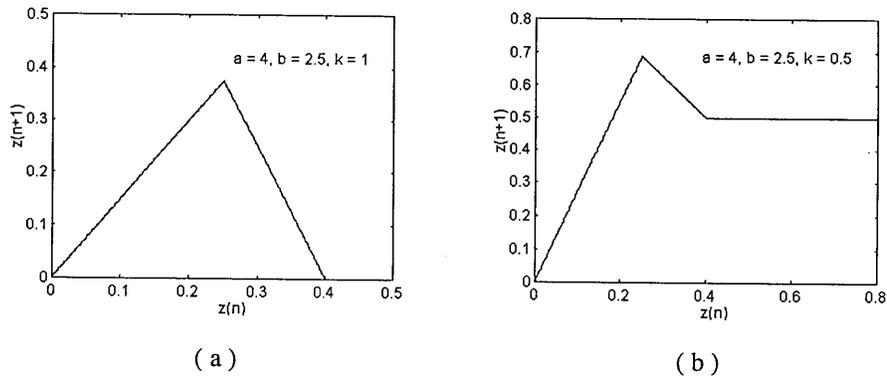, height=6cm}
\caption{The one-dimensional map representing neural pair dynamics for 
(a) $k = k^{\prime} = 1$ and (b) $k = k^{\prime} \neq 1$.
\label{fuinfig3} }
\end{center}
\end{figure}
The fixed points of this system are, $Z^*_1~=~0$ and $Z^*_2~=~1/(1+b)$.  
$Z^*_1$ is stable for $a-b~<~1$, whereas $Z^*_2$ exists only when $a-b~>~1$, 
and is stable for $b~<~1$. Beyond this, chaotic behavior is observed unless 
the maximum output value, i.e., $1-(b/a)$, iterates to the region where
$Z > 1/b$. 
The parameter space diagram is shown in Fig. 4. 

\begin{figure}[htbp]
   \vspace{0.5cm}
     \begin{center}
     \epsfig{file = 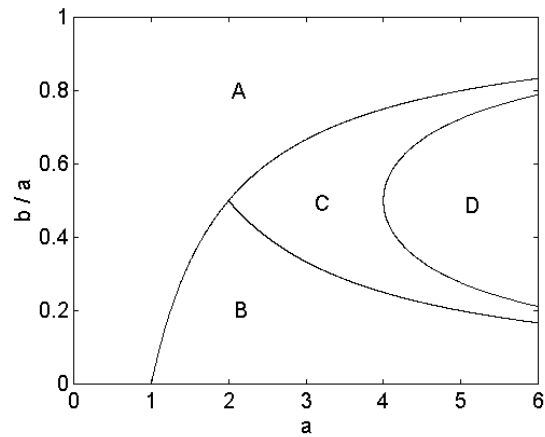, height=6cm}
\caption{The activation gain $a$ vs. $(b/a)$ parameter space for $k 
= k^{\prime} = 1$. 
Region A: $z^*=0$ stable, B: $z^*=1/(1+b)$ stable, C: chaos, D:  
coexistence of $z^*=0$ and a fractal chaotic invariant set. 
\label{fuinfig4} }
   \end{center}
\end{figure}
Along the line $b/a = 0.5$, 
we get the symmetric tent map scenario. So the Lyapunov exponent along 
this curve grows as $\lambda = {\log}_e (b)$ for $0<a<4$. 
This is one of the two special cases where an analytical expression for 
$\lambda$ can be obtained. The other instance is when the map's invariant 
probability distribution, $P(Z) = 1$. This occurs when  
\begin{equation} 
F~(1/a) = 1-(b/a) = 1/b. 
\end{equation} 
Along the curve defined by the above relation, the Lyapunov exponent evolves 
with the parameter $b/a$ according to 
\begin{equation} 
\lambda = -b/a~ {\log}_e (b/a) - (1 - (b/a)) {\log}_e (1 - (b/a)). 
\end{equation} 

\begin{figure}[htbp]
   \vspace{0.25cm}
      \begin{center}
        \epsfig{file = 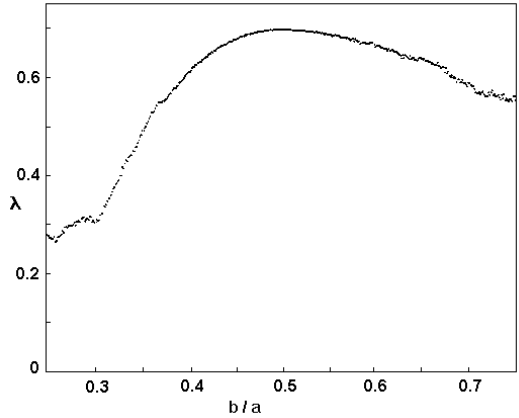, height=6cm}
\caption{Lyapunov exponent of the chaotic dynamics for $k = 
k^{\prime} = 1$ and $a = 4.0$. At $b/a = 0.5$, the entire interval $[0,1/b]$ 
is uniformly visited.
       \label{fuinfig5} }
    \end{center}
\end{figure}
In general, $\lambda$ has to be obtained computationally.  
Fig. 5 shows $\lambda$ 
plotted against $b/a$ for $a=4$, when the map is in the chaotic region.  
A sharp drop to zero is observed in both the terminal points, indicating 
sharp transition between chaotic and fixed-point behavior at $b/a=0.25$ and 
0.75. 
At $b/a = 0.5$, the entire interval $[0,1/b]$ is uniformly visited by the 
chaotic trajectory ($P(Z)=1$). This corresponds to ``fully-developed chaos'' 
in the symmetric tent map for which $\lambda = {\log}_e (2) \simeq 0.693.$  
 
When $F(1/a) > 1/b $, the interval $[0,1/b]$ is divided into a 
chaotic region of measure zero, defined on a non-uniform Cantor set (in 
general) and an ``escape set'' which maps to $Z^*_1 = 0$.  
This is because, for $Z \in (1/b(a-b), (b-1)/b^2)$, 
F(Z)=0. Any time an iterate of $Z$ falls in this region, in the next 
iterate the trajectory will converge to $Z^*_1$.  
The points left invariant after one iteration, will be in the 
two intervals $[0, 1/b(a-b)]$ and $[(b-1)/b^2, 1]$. The phase 
space is thus fragmented into two invariant regions. 
After $n$ iterations, there will be   
$2^n$ fragments of the chaotic invariant set, with $n!/r!(n-r)!$ ($r=0,1, 
\ldots, n$) intervals of length $(a-b)^{r} (1-b)^{r-n}$. The fragmentation of 
the phase space, 
therefore, has a {\em multifractal} nature  \cite{mccauley93}. 
 
The presence of 
multiple length scales is due to the fact that the slope magnitude of 
the map is not constant throughout the interval $[0, 1/b]$. 
It is to be noted that, even for $Z$ not belonging to the fractal 
invariant set, the trajectory 
might show long {\it chaotic transients} until at some iterate it maps to 
$Z^*=0$. 
For $b/a=0.5$, the map has a constant slope. As a result, 
the Cantor set is uniform, having exact geometrical self-similarity 
and a fractal dimension, $D = {\log}_e (2)/{\log}_e (b)$. So, the phase 
space of the coupled system has a fractal structure in this parameter 
region, i.e., where $1-(b/a) > 1/b $. 
 
\begin{figure}[htbp]
\vspace{1cm}
   \begin{center}
   \epsfig{file = 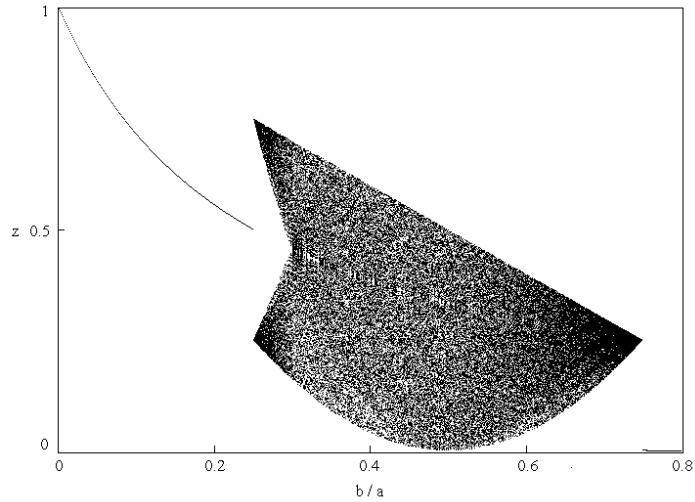, height=7cm}
   \caption{Bifurcation diagram for $k = k^{\prime} = 1$ at $a=4.0$.
     \label{fuinfig6} }
  \end{center}
\end{figure}
Fig. 6 shows the bifurcation structure of the map for $a=4$. For $b/a < 0.25$, 
the fixed point $Z^*_2$ is stable. At $b/a = 0.25$ it becomes unstable, 
leading to bands of chaotic behavior. 
The chaotic bands collide with the unstable 
fixed point $Z^*_2$ at $b/a \simeq 0.2985...$ and merge into a single 
chaotic band. 
This band-merging transition is an example of {\em crisis} \cite{goy83} and 
has been studied in detail for the symmetric tent map  
\cite{yoshida}. 
The $b$-value at which the band-merging occurs for a given value of $a$, 
can be obtained analytically by solving the quartic equation:  
\begin{equation} 
b^4 + (1-2a) b^3 + (a^2 - a) b^2 + a b + (a - a^2) = 0. 
\end{equation} 
For $2<a<2.5$, all the roots are complex, implying that band-merging does not 
occur over this range of $a$-values.  
 
Uniform chaotic behavior occurs at $b/a = 0.5$ (the entire interval 
[0,1/b] is uniformly visited by the chaotic trajectory). The chaotic 
band collides with the unstable $Z^*_2$ again at $b/a = 0.75$. This 
boundary crisis destroys chaos and stabilizes the fixed point $Z^*_1 = 0$. 

\subsection{$k = k^{\prime} \neq 1$}

This represents the condition when the connection weightages are such that, 
$W_{xy}/W_{xx}=W_{yy}/W_{yx} = k$, ($a  > b$).  
The dynamics is given by the following map (Fig. 3(b)) 
\begin{equation} 
\begin{array}{lll} 
Z_{n +1} & = & (a~-~kb)~Z_n, ~~{\rm if}~~0~\leq Z_n \leq~1/a,\\ 
        & = & 1~-~kb Z_n, ~~{\rm if}~~~ 1/a~< Z_n \leq~1/b,\\ 
        & = & 1~-~k, ~~{\rm otherwise}. 
\end{array} 
\end{equation} 
The key difference with the earlier case is that now the dynamics supports 
superstable period-$m$ orbits ($m \geq 2$). This is a result of the 
existence of a 
region of zero slope ($Z~>~1/b)$ giving a non-zero output.  
There are two fixed points of the map, $Z^*_1=0$ (as before), and, 
$$ 
\begin{array}{lll} 
Z^*_2 & = & 1-k,~ {\rm if} ~0<k<1-(1/b),~ {\rm or,}\\ 
      & = & 1/(1+kb),~ {\rm if}~ (a-1)/b>k>1-(1/b). 
\end{array} 
$$ 
$Z^*_2~=~1-k$, if it exists, is superstable, as the local slope is zero. 
On the other hand, $Z^*_2~=~1/(1+kb)$ is stable, only if $bk < 1$. 
If the fixed points are unstable, but iterates of $Z$ fall in the region 
$Z > 1/b$, superstable periodic cycles occur. The fixed point, $Z^*_1~=~0$, 
becomes stable when $(a - bk) < 1$. Chaotic behavior occurs if none 
of the fixed points are stable, and no iterate of $Z$ falls in the region 
$Z > 1/b$.   
The $(b/a)$ vs. $k$ parameter space diagram in Fig. 7 
(for $a=4$) shows the different dynamical regimes that are observed.  

\begin{figure}[htbp]
  \vspace{0.5cm}
     \begin{center}
     \epsfig{file = 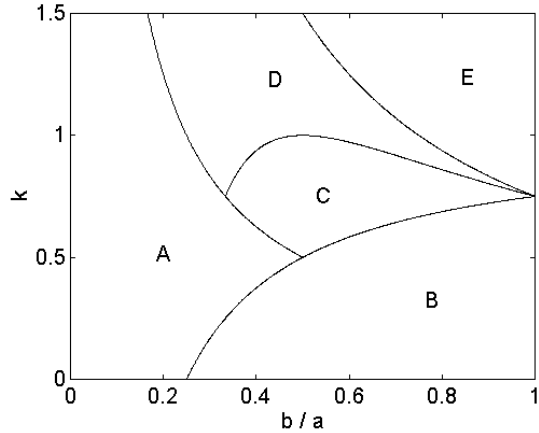, height=6cm}
       \caption{The $(b/a)$ vs. $k$ parameter space  
for $k = k^{\prime} \neq 1$ at $a=4.0$. Region A: $z^*=1/(1+kb)$
stable,  B: $z^*=1-k$ stable, C: superstable periodic cycles, 
D: chaos, E: $z^*=0$ stable. 
      \label{fuinfig7} }
  \end{center}
\end{figure}

\begin{figure}[htbp]
\vspace{1cm}
   \begin{center}
   \epsfig{file = 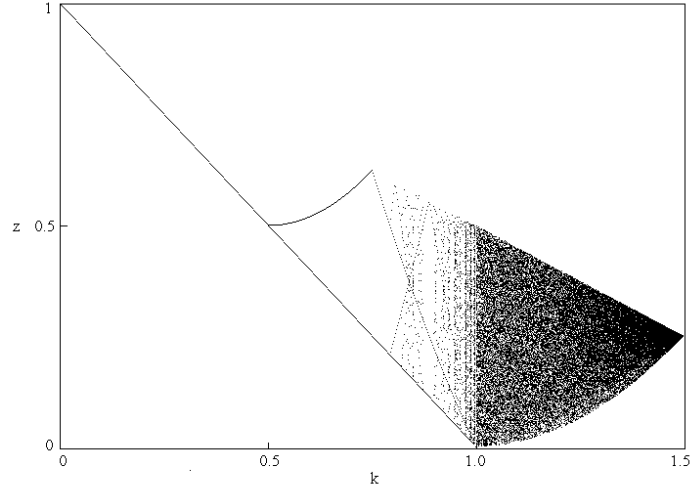, height=7cm}
\caption{Bifurcation diagram for $k = k^{\prime} \neq 1$ at $a=4.0, 
b=2.0$. }
\label{fuinfig8}
  \end{center}
\end{figure}

The bifurcation diagram for $a=4, b=2$ (Fig. 8) shows how the dynamics 
changes with $k$. For $ 0 \leq k < 0.5 $, $Z^*_2=1-k$, is the stable 
fixed point. At $k=0.5$, $Z^*_2$ becomes unstable, giving rise to a 
superstable period-2 cycle. A periodic regime is now observed, 
which was absent in the previous case. 
The periodic orbits initially follow a {\it period-doubling} sequence 
until a period-32 ($=2 \times 2^4$) orbit gives rise to a period-48 
($=3 \times 2^4$) one. This occurs as a result of border-collision 
bifurcations by which ``period-2 to period-3'' bifurcations have been 
seen to occur \cite{nusse92}. In the above instance, each of the sixteen 
period-2 orbits give rise to a period-3 orbit. The structure of the 
superstable periodic orbits is quite complex. The length of 
the cycles is plotted against $k$ in Fig. 9. 
\begin{figure}[htbp]
  \vspace{0.5cm}
    \begin{center}
      \epsfig{file = 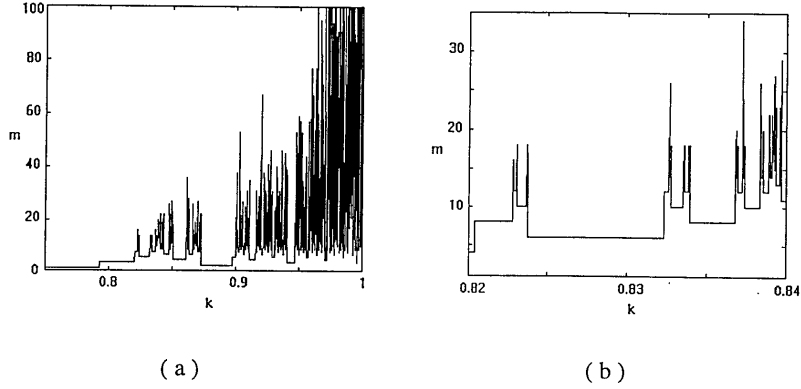, height=6cm}
\caption{Length of superstable periodic cycles, $m$, of the 
excitatory-inhibitory neural pair ($a=4$, $b =2$) for  
(a) 0.75 $\leq$ k $\leq$ 1, and (b) 0.82 $ \leq$ k $\leq$ 0.84. Note the
self-similar  structure of the intervals.
\label{fuinfig9} }
\end{center}
\end{figure}
The remarkable 
self-similar structure of the intervals is to be noted. Numerical 
studies indicate that cycles of all periods exist having the 
following ordering: between any superstable 
period-$m$ and period-$(m+1)$ cycle, there exists an interval of $k$ 
for which a period-$(m+2)$ orbit is superstable. 
At $k=1.0$ all periodic orbits become unstable, leading to 
onset of chaos. The chaotic behavior persists till $k=1.5$, 
when $Z^*_1 = 0$ becomes stable. 
 
Let us now study the effect of introducing a {\em threshold}. A positive 
threshold, $t_0>0$, applied to both $F_a$ and $F_b$,  
has the effect of shifting the map to the right by $t_0$. This allows 
the segmentation of activation state $(X,Y)$-space, according to dynamical 
behavior. For initial conditions lying in the region bounded by the two 
straight lines, $Y=(X-t_0)/k$ and $Y=(X-t_0)/(k - 1/ak)$, the trajectories 
are chaotic, provided the maximum point of the map, $F(Z) = 1 - (kb/a)$, 
does not iterate into the region $Z>t_0+(1/b)$. 
For the region, $Y>(X-t_0)/k$, any iterate will map to the fixed 
point, $Z^*=0$. Initial conditions from $Y<(X-t_0)/(k - 1/ak)$ 
will map to the chaotic region, 
if the maximum point of the map does not iterate into $Z>t_0+(1/b)$. 
Otherwise, a fractal set of initial conditions will give rise to bounded 
chaotic motion, the remaining region falling in  
the ``escape set'', eventually leading to periodic orbits. 
 
A negative threshold shifts the map to the left. This has the consequence 
of destabilizing the fixed point $Z^*=0$. If the other fixed point is also 
unstable then this will imply that all possible $(X,Y)$ values will 
give rise to periodic or chaotic trajectories. 
This indicates the possible existence of a global 
chaotic attractor. 

\subsection{$k \neq k^{\prime}$}

This corresponds to the condition when all the connection weights are 
different. The dynamics is irreducible to 1-dimension. We need to consider 
only the positive $(Z,Z^{\prime})$ region, as otherwise, $(0,0)$ 
is the stable fixed point. In the non-zero region, different dynamical 
behavior may occur depending on the region where the fixed point occurs 
and on its stability. One of the fixed points is $(Z,Z^{\prime})=(0,0)$, 
whose stability is determined by  
obtaining the eigenvalues of the corresponding Jacobian, 
$$ 
{\bf J} = \begin{array} {|cc|} 
a & -kb\\ 
a & -k^{\prime} b 
\end{array} ~~~~. 
$$ 
Evaluating the above matrix, gives the following condition 
\begin{equation} 
-2 < (a-k^{\prime} b) \pm [(a-k^{\prime}b)^2 - 
4ab(k-k^{\prime}) ]^{1/2} < 2, 
\end{equation} 
for stability of the fixed point. 
 
The other fixed point may occur in any one of the four following regions 
of the $(Z,Z^{\prime})$-space: 
  
{\bf Region I}: $0<Z<1/a$, $0<Z^{\prime}<1/b$.\\ 
$(Z,Z^{\prime})=(0,0)$ is the only fixed point.  
 
{\bf Region II}: $0<Z<1/a$, $Z^{\prime}>1/b$.\\ 
The fixed point is $(Z,Z^{\prime}) = (k/(a-1), 
(a(k-k^{\prime})+k^{\prime})/(a-1))$, which is stable if $ -1< a <1 $. 
 
{\bf Region III}: $Z>1/a$, $0<Z^{\prime}<1/b$.\\ 
The fixed point is $(Z,Z^{\prime})=((1+b(k^{\prime}-k))/(1+bk^{\prime}), 
1/(1+bk^{\prime}))$, which is stable if $-1<k^{\prime}b<1$. 
 
{\bf Region IV}: $Z>1/a$, $Z^{\prime}>1/b$.\\ 
The fixed point is $(Z,Z^{\prime})=(1-k,1-k^{\prime})$. This is a superstable 
root, as the local slope is zero under all conditions. 
 
\begin{figure}[htbp]
  \vspace{0.5cm}
     \begin{center}
       \epsfig{file = 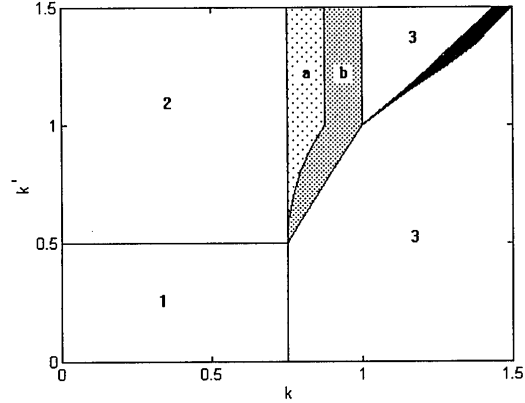, height=6cm}
\caption{The $(k, k^{\prime})$ parameter space for $a=4.0$ and $b=2.0$. 
Region 1: $x^*$=1, $y^*$=1 stable, 2: $x^*$=1, $y$ has period-2 cycles, 
3: $x^*$=0, $y^*$=0 stable, $a$: Both $x$ and $y$ have period-2 
cycles, $b$: $x$ and $y$ show period-$m$ cycles ($m > 2$). Fully chaotic 
behavior occurs in the dark wedge-shaped region in 3. In addition, 
fractal intervals showing chaos occur in region $b$.
\label{fuinfig10} }
\end{center}
\end{figure}

The abundance of tunable parameters in this case makes detailed simulation 
study extremely difficult. However, some preliminary studies in the $(k, 
k^{\prime})$ parameter space (keeping the other parameters fixed) gives 
indication of dynamics similar to that seen in cases $(i)$ and $(ii)$. 
The $(k, k^{\prime})$ parameter space is shown in Fig.~\ref{fuinfig10}
for $a=4$,
$b=2$.
  
A variety of dynamical behaviors is observed - from fixed points to periodic 
cycles to chaos, as indicated by the different regions. In addition, 
there are regions exhibiting periodic behavior which have 
fractal intervals of chaotic activity embedded within them. 

\section{Extension to Large Networks} 

In the previous section, the behavior of a pair of excitatory-inhibitory 
neurons (number of neurons, $N=2$) was shown to have sufficient complexity. 
The dynamics of a N-neuron network ($N>>2$) described by  
$$ 
{\bf X_{n+1}} = F(\sum {\bf W}.{\bf X_{n})}, 
$$ 
where ${\bf X_n}$ is the set of $N$ activation state values (both 
excitatory and 
inhibitory neurons), and ${\bf W}$ is the matrix of connection weights 
between different neurons. The full range of behavior shown by such a system 
will be impossible to 
study in detail, as the number of available tunable parameters are 
too large to handle. However, under certain restrictions, the dynamics 
of such large networks can be inferred. 
 
Let $W_{ij}$ denote the connection weight from $j$th to the $i$th neuron. 
Then, under the condition 
\begin{equation} 
W_{i~j+1}/W_{i~1}=k_j, ~~~~(k_j = {\rm  
constant ~~for ~~a ~~given} ~~j=1, \ldots, N), 
\end{equation} 
the N-neuron network dynamics is reducible to that of  
a 1-dimensional map with $(N+1)$ linear segments (for $t=0$). 
The occurrence of ``folds'' in a map have already been shown to be 
responsible for creation and persistence of localized coherent structures  
within a chaotic flow \cite{shinbrot}. As in this case, the resultant map 
will have a number of such folds, the system might 
show coexistence of multiple chaotic attractors (isolated from each other). 
A simple example to illustrate this point is a fully connected network of 
four neurons: two excitatory ($x_1$,$x_2$) 
and two inhibitory ($y_1$,$y_2$). Let $a_i$ and $b_i$ 
represent the slope of the transfer functions for the $i$th excitatory 
and inhibitory neurons, respectively. The 4-dimensional dynamics 
is reducible to the 1-dimensional dynamics of $z=x_1 - k_1 y_1 + k_2 x_2 
- k_3 y_2$. Simulations were carried out for the set of parameter 
values: $(a_1=4.6, a_2=4.0, b_1=3.6, b_2=1.6)$ and $(k_1=0.7, k_2=1.0, 
k_3=1.1)$. Furthermore, $x_2$,$y_2$ have a threshold equal to $1/b_1$. 
Fig. 11(a) shows the return map and time evolution of $z$ in the absence 
of any bias. There is only a single global chaotic attractor in this case. 
When a small negative bias is applied to the whole network, the previous 
attractor splits into two coexisting isolated attractors having  
localized chaotic activity. Which attractor the system will be in, 
depends upon the initial value it starts from.
\begin{figure}[htbp]
   \vspace{0.5cm}
     \begin{center}
      \epsfig{file = 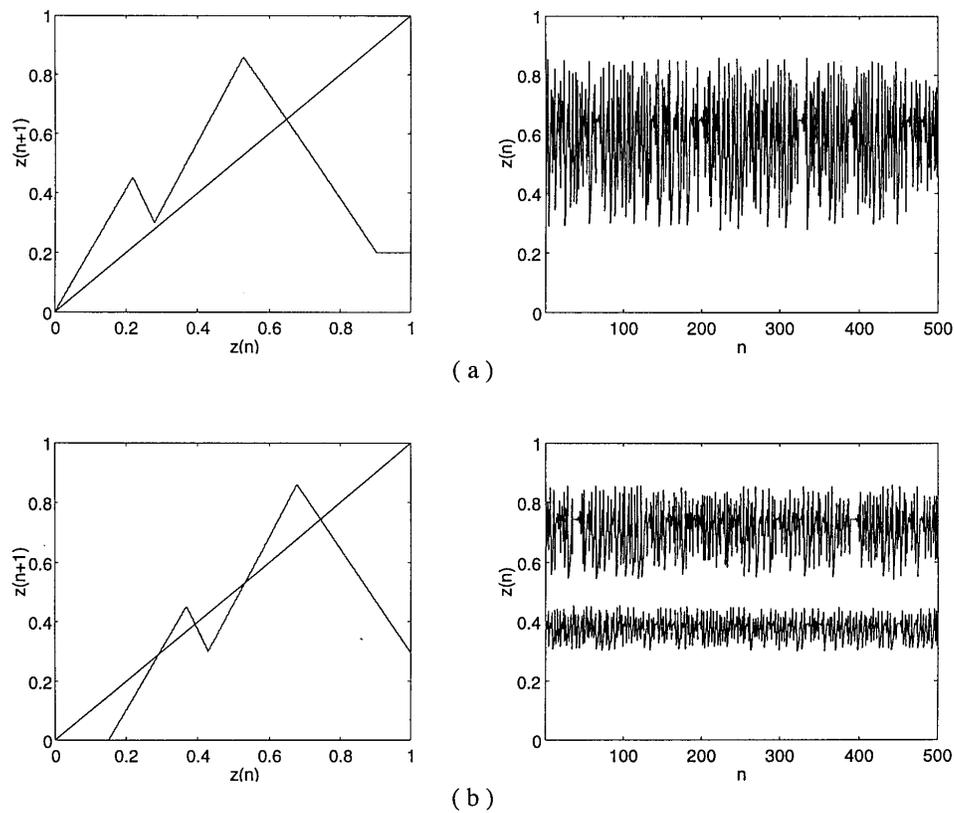, height=12cm}
\caption{The return map and time evolution of the reduced 
variable, $z$, for a 4-neuron network (for details see text), 
with (a) bias = 0 and (b) bias = -- 0.15. In the former there 
is a single global chaotic attractor. For non-zero bias, there are two 
co-existing chaotic attractors. Time evolution of $z$ starting from 
two initial conditions belonging  to different attractors are superposed.
\label{fuinfig11} }
\end{center}
\end{figure}
Fig. 11(b) shows the return map for a bias value of -0.15 and the  
superposed time evolutions of 
$z$ starting from initial conditions belonging to two different attractors. 
So, an increase in bias, can cause transition from global chaos to localized 
chaotic regions. 
 
This property can be used to simulate a proposed mechanism of olfactory 
information processing \cite{freeman92}. It has been 
suggested that the olfactory system maintains a global attractor with 
multiple ``wings'', each corresponding to a specific class of odorant. 
During each inhalation, the system moves from the central chaotic repeller 
to one of the wings, if the input contains a known stimulus. The continual 
shift from one wing to another via the central repelling zone has been 
termed as chaotic ``itinerancy''. This forms the basis of several chaotic 
associative memory models. 
 
The above picture can be observed in the 
present model by noting that, if the external input has the 
effect of momentarily increasing the bias from a negative 
value to zero, then the 
isolated chaotic regions merge together into a single global attractor. 
In this condition, the entire region is accessible to any input state. 
However, as the bias goes back to a small negative value, the different 
isolated chaotic attractors re-emerge, and the system dynamics is 
constrained into one of these. Sustained external stimuli will cause 
the gain parameters to decrease (adaptation), thereby decreasing the 
local slope of the map. If the 
stimulus is maintained, the unstable fixed point in the isolated region 
will become stable leading to a fixed-point or periodic behavior. 
The above scenario, in fact, is the basis of using the proposed model 
as an associative memory network. 
 
\section{Information Processing with Chaos} 

Chaotic dynamics enables the microscopic sensory input 
received by the brain to control the macroscopic activity that constitutes 
its output. This occurs as a result of the selective sensitivity of chaotic 
systems to small fluctuations in the environment and their 
capacity for rapid state transitions. 
On the other hand, chaotic attractors are globally extremely robust. 
These properties indicate that the utilization of chaos by biological 
systems for information processing can indeed be advantageous. 
It has been suggested, based on investigations into cellular automata, 
that complex computational capabilities emerge at the ``edge of chaos'' 
\cite{langton}.  
 
Based on this notion, efforts are on to use chaos in neural network models 
to achieve human-like information processing capabilities. 
Chaotic neural networks have been already been applied in designing  
associative memory networks \cite{ishii96} 
and solving combinatorial optimization problems, using chaos to carry 
out an effective stochastic search \cite{inoue92}. 
The superposition of chaotic maps for information processing has also been 
suggested before \cite{andrey97}. 
 
The model presented 
in this paper can be used for a variety of purposes, classified as follows: 
 
{\bf Associative memory}:  
A set of patterns (i.e., specific network state configurations) 
are stored in the network as attractors of the system 
dynamics, such that, whenever a distorted version of one of the 
patterns is presented to the network as input, 
the original is retrieved upon iteration. The distortion has to be small 
enough so that the input pattern is not outside the basin of attraction 
of the desired attractor. In networks using convergent dynamics, the stored 
patterns necessarily have to be time-invariant or at most, periodic. 
 
Chaos provides rapid and unbiased access to all attractors, any of which 
may be selected on presentation of a stimulus, depending upon the network 
state and external environment. It also acts as a ``novelty detector'', 
classifying a stimulus as being previously unknown, by not converging 
to any of the existing attractors.  
This suggests the use of chaotic networks for {\it auto association}. 
 
In the previous section, the basic mechanism for constructing an 
associative memory network has been described. 
In this proposed model, both constant and periodic sequences can be stored.  
This is made possible by introducing ``folds'' in the return map of the 
network, so that a large number of isolated regions are produced. The 
nature of the dynamics in a region can be controlled by altering 
the gain parameters of individual neurons. Accessibility to a given 
attractor depends upon the initial condition of the network and the 
input stimulus. So, regions with fixed-point or periodic attractors 
may be embedded within regions having chaotic behavior. 
In addition, chaotic trajectories confined within a specific region 
can also be generated when presented with a short-duration 
input stimulus belonging to that region. 
``Novelty detection'' is implemented in the above model by making the basins 
of attractors (corresponding to the stored patterns) 
of some pre-specified size. 
Input belonging outside the region, therefore, cannot enter the basin 
and will not be able to converge to the stored pattern. 
 
{\bf Pattern classification}: 
In this information processing task, different input sets need 
to be classified into a fixed number of categories.  
Decision boundaries, i.e., boundaries between the different classes are 
constructed by a ``training session'' where the network is presented 
with a series of inputs and the corresponding class to which they belong. 
In the proposed model, classification can be on the basis of dynamical 
behavior. For example, input sets belonging to different input classes 
may give rise to different periodic sequences. Otherwise, the distinction 
can be made between categories of inputs which give rise to chaotic and 
non-chaotic trajectories. For a pair of  
neurons ($N=2$), under the condition $k = k^{\prime}$, 
linear separation of the $(X,Y)$-space can be done (as shown 
above). By varying the parameters $k$ and $b$, the orientation and size 
of the class regions can be controlled. If $k \neq k^{\prime}$ and $N>2$, 
nonlinear decision boundaries between different classes can be generated. 
By using suitably adjusted weights, any arbitrary classification can be 
achieved. 
 
\begin{figure}[htbp]
  \vspace{0.5cm}
    \begin{center}
      \epsfig{file = 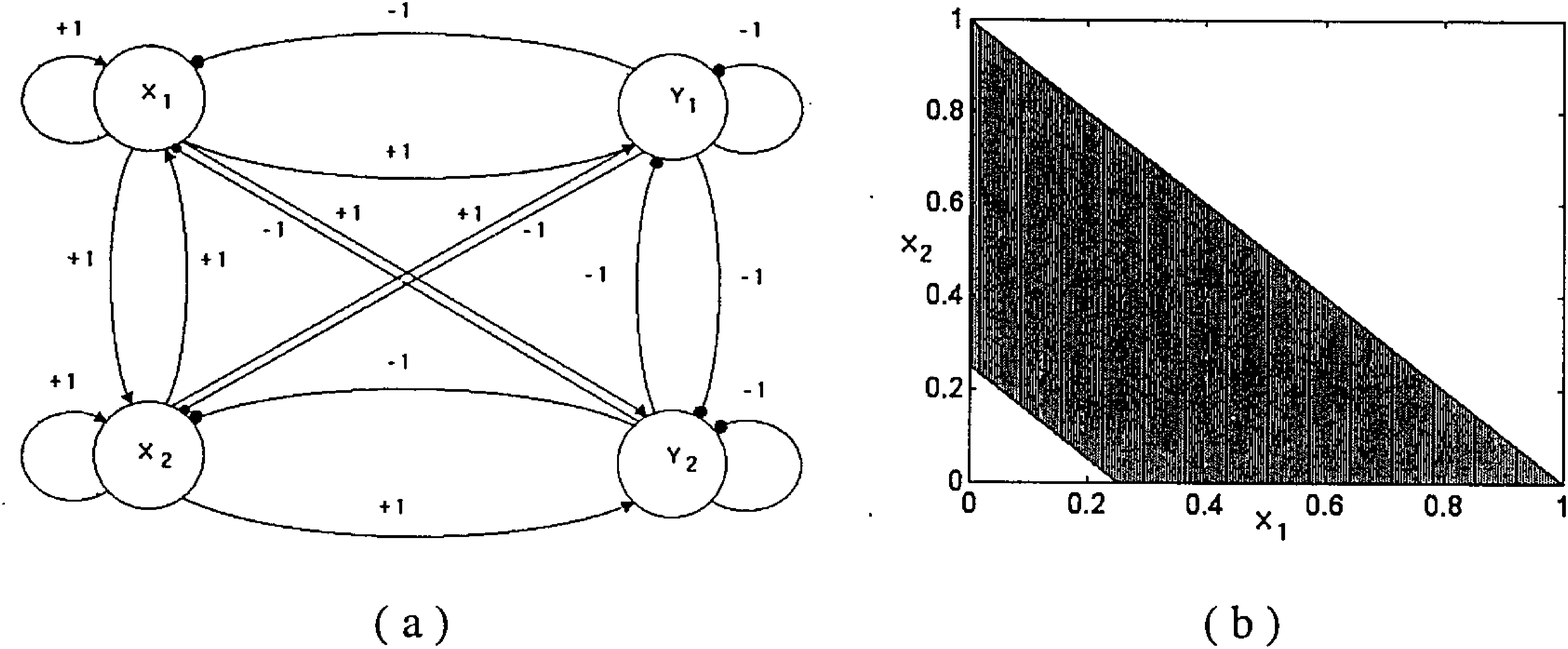, height=6cm}
\caption{(a) A fully connected 4-neuron network with  
2 excitatory ($x_{1,2}$) and 2 inhibitory ($y_{1,2}$) neurons. The arrows 
and circles represent excitatory and inhibitory synapses, respectively. 
(b) The $(x_1,x_2)$ phase space shows the basin of the chaotic attractor 
(shaded region) for threshold, $t$=0.25. The unshaded region 
corresponds to fixed point behavior 
of the network. 
\label{fuinfig12} }
\end{center}
\end{figure}
An example of nonlinear decision boundary generation is shown in Fig. 12. 
The network used for this purpose consists of 4 neurons - 2 excitatory 
($x_1, x_2$) and 2 inhibitory ($y_1, y_2$) (fig. 12(a)). The gain 
parameters are $a_{x_1,x_2}=2$ and $b_{y_1,y_2} =1$. All the 
neurons have a threshold, $t$. The network is fully connected 
with all weightages equal to unity. The input stimulus is taken to be 
the initial value of the excitatory neurons and the inhibitory neuronal 
states are initially taken to be zero. 
 
As shown in fig. 12(b), for $t$ = 0.25, the $(x_1, x_2)$ phase space 
is segmented into basins leading to either fixed point or 
chaotic attractors. By increasing $t$, the width of the chaotic 
band can be reduced. Changing the initial value of the inhibitory 
neurons will cause translation of the band and manipulating the connection 
weights gives a rotation to the band. Thus, any general transformation 
can be applied to the segment. More complex network connections might 
permit segmenting isolated box-like regions in the phase space. 
This possibility is currently under investigation.  
 
{\bf System Dynamics Approximation}: 
A system may be described by a nonlinear input-output relation, 
$$ 
Y={\bf G}(X), 
$$ 
where the mapping function, ${\bf G}$, is unknown. 
By having access to a limited set of 
input-output pairs, the function has to be approximated - in effect, 
building a system simulator. In the present model, a sufficient 
number of coupled neurons can be used to construct any arbitrary 
piecewise linear input-output relation. By use of a suitable learning 
rule, the available data set can be used to determine the gain 
parameters, thresholds and connection weights of the network. A close 
approximation of the system dynamics will enable prediction and control 
of its behavior. The approximation's accuracy is not restricted to 
systems with piecewise linear functions - but can also give good 
qualitative reconstruction of smooth nonlinear systems. 
 
{\bf Periodic sequence generation}: 
Capability for periodic sequence generation can be exploited for modeling 
{\it central pattern generators}. These are a class of biological neural 
ensembles which control well-defined rhythmic muscle movements such as  
swimming, running, walking, breathing, etc. Usually they are found in 
the spinal cord, producing periodic sequences without feedback from 
the motor system or higher-level control. The ability to generate multiple 
sequences from the same neural assembly is another interesting 
feature. Postulating the existence of 
single pacemaker neurons acting as the `system clock' to initiate 
periodic activity cannot explain all the observed phenomena. 
The existing network models for simulating this 
behavior mostly suffer from the drawback that they cannot generate 
multiple non-overlapping sequences. This shortcoming is overcome in the 
model presented here. For $N=2$ and $k=k^{\prime} \neq 1$, 
a rich variety of periodic sequences can 
be chosen from the same network, simply by altering the gain parameters 
by a very small amount. As mentioned above, numerical investigations 
indicate that cycles of any period can be generated by suitably 
altering the value of $k$.

\section{Discussion and Future Work}

The network has been studied in detail only for $N=2$. Nonetheless it 
shows capability for supporting extremely complex behavior. Under 
certain restrictions, the dynamics for $N>>2$ networks can also be 
understood. Relaxation of these restrictions will provide a challenging task 
for the future. 
 
One important point not touched here is the issue of {\it learning}. The 
connection weights \{$W_{ij}$\} have been assumed constant, as they change 
at a much slower time scale compared to that of the activation state 
values. However, modification of the weights due to learning will 
also cause changes in the dynamics. Such bifurcation behavior, induced 
by weight changes, will have to be taken into account when devising 
learning rules for specific purposes. The interaction of chaotic 
activation dynamics at a fast time scale and learning dynamics  
on a slower time scale might yield richer behavior than that seen in the 
present model.  
An interesting study for the future will be to introduce time-varying 
Hebbian synaptic weights, evolving according to 
$$ 
W_{ij} (n+1) = W_{ij} (n) + \epsilon X_i (n) X_j (n), 
$$ 
where $X(n)$ and $W(n)$ denote the neuron state and connection weight at the 
$n$th instant, and $\epsilon$ is related to the time-scale of the synaptic 
dynamics. The combined effect of such synaptic dynamics with 
the chaotic network dynamics (for small $N$) described above 
might lead to significant departure from the overall features 
described here. 
 
Real biological systems reside in an extremely noisy environment. This 
is incorporated into neural models by using stochastic updating rule 
and/or  explicit introduction of a term representing external noise. We plan 
to introduce similar features in the proposed model. In dissipative 
chaotic systems, the effect of external noise seems to be limited to 
destroying the fine 
structure of the bifurcation sequence \cite{crutch}. The interaction 
of deterministic chaos and stochastic noise in the network 
will be interesting to study. 
 
Another possible modification to the present model will be to incorporate 
time-varying operating conditions. In \cite{lwang96} time-dependence 
of a suitable system parameter was shown to give rise to interesting 
dynamical behaviors, e.g., transition between periodic oscillations and 
chaos. This suggests that varying the environment can facilitate memory 
retrieval if dynamic states are used for storing information in a neural 
network. In the proposed network, periodic variations can 
be incorporated into the available system parameters: gain, threshold and 
connection weights. In nonlinear systems, time-varying dynamics often leads 
to the phenomenon of ``stochastic resonance'' where a subthreshold signal 
is amplified by the background noise. It has been observed that replacing 
noise by chaos also gives rise to this resonance behavior in 
a bimodal piecewise linear map \cite{sinhachak}. 
It will be of interest to see whether such phenomena can occur in the 
present model, on the introduction of temporal variation in the parameters. 
The occurrence of ``resonance''  
might enhance the capability of the network to detect subthreshold signals, 
through signal amplification by deterministic `noise'.

\end{document}